\documentclass[aps,reprint,prd,nofootinbib,noshowkeys,longbibliography,floatfix,superscriptaddress,preprintnumbers]{revtex4-1}
\usepackage[utf8]{inputenc} 
\usepackage{hyperref} 
\usepackage{amsmath} 
\usepackage{amsfonts}
\usepackage{amsthm}
\usepackage{amssymb}
\usepackage{graphicx} 
\usepackage{color} 
\usepackage[sort&compress]{natbib}
\usepackage[normalem]{ulem}

\DeclareMathOperator{\tr}{Tr}

\begin{document}

\title{Instanton solution for Schwinger production of ’t Hooft-Polyakov monopoles}
\date{March 23, 2021}

\author{David L.-J. Ho}
\email{d.ho17@imperial.ac.uk}
\author{Arttu Rajantie}
\email{a.rajantie@imperial.ac.uk}
\affiliation{Department of Physics, Imperial College London, SW7 2AZ, UK}
\preprint{IMPERIAL-TP-2021-DH-04}

\begin{abstract}
We present the results of an explicit numerical computation of a novel instanton in Georgi-Glashow SU(2) theory. The instanton is physically relevant as a mediator of Schwinger production of 't Hooft--Polyakov magnetic monopoles from strong magnetic fields. In weak fields, the pair production rate has previously been computed using the worldline approximation, which breaks down in strong fields due to the effects of finite monopole size. Using lattice field theory we have overcome this limit, including finite monopole size effects to all orders. We demonstrate that a full consideration of the internal monopole structure results in an enhancement to the pair production rate, and confirm earlier results that monopole production becomes classical at the Ambj{\o}rn-Olesen critical field strength.
\end{abstract}

\maketitle

\section{Introduction}

Magnetic monopoles are hypothetical particles consisting of a single, isolated magnetic pole. Though commonly omitted from Maxwell's equations, there are no known theoretical barriers to their existence, and they are predicted by a wide range of theories extending the Standard Model. A recent review of the theoretical and experimental status of magnetic monopoles can be found in Ref.~\cite{mavromatos2020magnetic}.

Magnetic monopoles can be included in a theory as elementary particles \cite{dirac1931quantised, cabibbo1962quantum, schwinger1966magnetic, zwanziger1970local, blagojevic1985quantum}, or can appear as solitonic excitations in a wide class of nonabelian gauge theories \cite{thooft1974magnetic, polyakov1974particle}. This work focuses primarily on solitonic monopoles, though the issue of finite size corrections is still present for elementary monopoles due to quantum effects \cite{goebel1970spatial, goldhaber2017magnetic}. In some models, the existence of dualities \cite{montonen1977magnetic} means that the distinction between elementary and solitonic excitations is not well defined: in such theories, our results may also apply to elementary particles.

In order to use the negative results of past experimental monopole searches to constrain the properties of the Universe, it is vitally important to gain a theoretical understanding of the mechanisms by which monopoles may be produced. Perturbative calculations, however, are doomed by the Dirac quantisation condition~\cite{dirac1931quantised}:
\begin{equation}
    e g = 2 \pi n,
\end{equation}
where \(e\) and \(g\) denote the quanta of electric and magnetic charge respectively, and \(n\) is an integer. The perturbative nature of the electric coupling means that magnetodynamics is necessarily nonperturbative. This means that it is not possible to compute the monopole production cross section in collisions of elementary particles using existing methods. Furthermore, it has been argued~\cite{witten1979baryons, drukier1982monopole} that the production of solitonic monopoles in proton-proton collisions is suppressed by a factor of \(\sim \mathrm{e}^{-4 / \alpha} \approx 10^{-236}\). This is due to the fact that solitons may be thought of as a coherent state of many elementary particles: with a small number of degrees of freedom in the initial state, the formation of a final state with many degrees of freedom is vanishingly unlikely.

A process of monopole-antimonopole pair production that circumvents both this suppression and the need for a perturbative expansion in the monopole coupling is the Schwinger effect~\cite{Sauter:1931zz, Schwinger:1951nm}. This is, at least in weak fields, a quantum mechanical process, by which a field is unstable to production of charged particle-antiparticle pairs. There are many methods of calculating the Schwinger production rate \(\Gamma\), but the method best applicable to strongly coupled particles is the instanton method~\cite{affleck1981monopole, affleck1981pair}, which computes the rate semiclassically. In weak fields, and when monopoles may be modelled as point particles, the result for magnetic monopoles is
\begin{equation}
    \Gamma_{\mathrm{m} \bar{\mathrm{m}}} \propto \exp \left(-\frac{\pi M^2}{g B} + \frac{g^2}{4} \right),
\end{equation}
where \(M\) is the monopole mass, and \(B\) is the strength of the (constant) external field. This is valid in the semiclassical limit
\begin{equation} \label{eq:semiclassicalLimit}
    \frac{g^3 B}{4 \pi M^2} \ll 1,
\end{equation}
and is accurate to all orders in the magnetic charge \(g\). The semiclassical limit can equivalently be viewed as the stipulation that the external field must be weak in units set by the monopole mass.

In this paper, we have computed the instanton solution for 't~Hooft-Polyakov monopole pair production in the Georgi-Glashow SU(2) theory at all relevant field strengths, thereby extending the calculation of Refs.~\cite{affleck1981monopole, affleck1981pair} beyond the worldline approximation, which is only valid in the weak field limit. We confirm our earlier result~\cite{ho2019classical} that monopole production becomes a classical process at the Ambj{\o}rn-Olesen critical field strength~\cite{ambjorn1988antiscreening}
\begin{equation}
    B_\mathrm{crit} = \frac{m_\mathrm{v}^2}{e},
\end{equation}
where \(m_\mathrm{v}\) is the charged vector boson mass, and \(e\) is the gauge coupling. Furthermore, we show that, in constant fields, the higher order corrections appear to universally enhance the exponential dependence of the monopole production rate.

The structure of this paper is as follows: in Section \ref{sec:theory} we outline the theoretical background behind our calculation, including the worldline instanton method and Georgi-Glashow SU(2) theory. In Section \ref{sec:methods} we present the lattice formulation of the theory and the numerical methods used to find the instanton solutions. In Section \ref{sec:results} we give our results, and in Section \ref{sec:conclusion} we conclude our arguments, with particular focus on the implications for deriving monopole mass bounds from heavy-ion collision.

\section{Theory} \label{sec:theory}

\subsection{Instantons for Schwinger production}
Instanton methods were first applied to Schwinger production in Refs.~\cite{affleck1981monopole, affleck1981pair} to calculate the production rate of strongly coupled particles in weak fields. While the computation of the Schwinger production rate is possible by many methods, worldline instantons---which are applicable when the produced particles can be considered pointlike---are particularly useful because they can be used for strongly coupled particles, inhomogeneous external fields~\cite{dunne2005worldline, dunne2006worldline, gould2018worldline} and finite temperatures~\cite{gould2017thermal}.

The starting point for the calculation, both for point particles and solitons, is the expression for the pair production rate
\begin{equation}
    \mathcal{V} \Gamma = -2 \mathop{\mathrm{Im}} \log \int \mathcal{D} [\phi] \, \mathrm{e}^{-S_\mathrm{E}}
\end{equation}

where \(S_\mathrm{E}\) is the Euclidean action of the theory (including the external field) \(\mathcal{V}\) is a spacetime volume, and \(\mathcal{D}[\phi]\) is used to denote the path integral over all fields in the theory. This may be approximated using the method of stationary phase; the dominant contribution to this path integral is from the lowest lying stationary point of the action that gives the action an imaginary part. This is a saddle point solution to the equations of motion with a single negative mode---an instanton. Using the dilute instanton gas approximation, one finds that
\begin{equation}
    \Gamma \propto \exp(-S_\mathrm{inst}),
\end{equation}
where \(S_\mathrm{inst}\) is the Euclidean action evaluated on the instanton solution to the equations of motion.

The approach of the worldline instanton method is to formally re-express the path integral over fields as an integral over all charged particle worldlines. This may be done exactly for scalar or spinor QED, and under certain approximations becomes analytically tractable. It can therefore be used to calculate the Schwinger production rate for point monopoles, or for 't Hooft--Polyakov monopoles in the limit where the monopole size is small compared to the instanton size.

In a constant external field the worldline instanton solution may be found analytically: the worldline path is circular with radius
\begin{equation} \label{eq:worldlineInstantonRadius}
    r_\mathrm{inst} = \frac{M}{g B}.
\end{equation}
The corresponding action is
\begin{equation} \label{eq:actionPointlikeLimit}
    S_\mathrm{inst} = \frac{\pi M^2}{g B} - \frac{g^2}{4}.
\end{equation}
The negative mode is a ``breathing'' mode, increasing or decreasing the instanton radius.

The worldline instanton solution is valid providing the monopole size is small compared to the size of the instanton. The classical monopole radius, in the case of both pointlike and solitonic monopoles, is
\begin{equation}
    r_\mathrm{m} \sim \frac{g^2}{4 \pi M}.
\end{equation}
This suggests that the small monopole condition \(r_\mathrm{m} \ll r_\mathrm{inst}\) is equivalent to the semiclassical limit \eqref{eq:semiclassicalLimit}. In fact, we will show in Section~\ref{sec:results} that, for `t Hooft--Polyakov monopoles, finite size effects begin to become apparent somewhat before semiclassicality breaks down.

\subsection{Georgi-Glashow SU(2) Theory}
The purpose of this work is to overcome the limitations of the worldline approximation by computing the instanton solution numerically, taking the internal structure of the monopole into account. In order to achieve this we must specify a field theory admitting finite energy magnetic monopole configurations. We choose the canonical example of Georgi-Glashow SU(2) theory~\cite{georgi1972unified}, which may be embedded in any Grand Unified Theory that contains the Standard Model.

The theory consists of an SU(2) gauge field \(A_\mu\) with an adjoint scalar field \(\Phi\): the four dimensional Euclidean Lagrangian is
\begin{equation} \label{eq:georgiGlashowLagrangian}
    \mathcal{L} = \frac{1}{2} \tr F_{\mu \nu} F^{\mu \nu} + \tr D_\mu \Phi D^\mu \Phi + V(\Phi),
\end{equation}
where
\begin{align}
    D_\mu \Phi^a &= \partial_\mu \Phi^a + i e \varepsilon^{abc} A_{\mu}^b \Phi^c, \\
    F_{\mu \nu}^a &= \partial_\mu A_\nu^a - \partial_\nu A_\mu^a + i e \varepsilon^{abc} A_\mu^b A_\nu^c, \\
    V(\Phi) &= \lambda\left(\tr(\Phi^2) - v^2\right)^2.
\end{align}
Here \(\mu, \nu = 1, 2, 3, 4\) are the Euclidean spacetime indices, and \(a,b,c = 1, 2, 3\) are SU(2) Lie algebra indices. The theory has two dimensionless parameters: the gauge coupling \(e\) and the scalar field self-coupling \(\lambda\), and the scalar field vacuum expectation value (vev) \(\sqrt{2}v\), which sets the scale.

The theory admits `t Hooft--Polyakov monopole solutions~\cite{thooft1974magnetic, polyakov1974particle} with magnetic charge
\begin{equation} \label{eq:monopoleCharge}
    g = \frac{4 \pi}{e},
\end{equation}
i.e. two units of Dirac charge quanta. The classical mass of this monopole is
\begin{equation} \label{eq:monopoleMass}
    M = \frac{4 \pi m_\mathrm{v}}{e^2} f(\beta),
\end{equation}
defining the boson mass ratio \(\beta = m_\mathrm{s} / m_\mathrm{v}\). For all values of \(\beta\), \(f(\beta) \sim 1\)~\cite{forgacs2005numerical}. The theory also admits higher charge monopoles and dyons~\cite{julia1975poles}, the latter being responsible for subleading contributions to the Schwinger production rate. In this analysis we focus only on the lightest monopole excitations, without electric charge: these are responsible for the dominant contribution to the Schwinger effect.

It is interesting to note that substitution of the `t Hooft--Polyakov monopole mass and charge into the weak field condition \eqref{eq:semiclassicalLimit} at the point of equality gives
\begin{equation}
    B = \frac{m_\mathrm{v}^2 f(\beta)^2}{e}.
\end{equation}
This is, to within an \(O(1)\) constant, equal to the Ambj{\o}rn-Olesen critical field strength~\cite{ambjorn1988antiscreening, ambjorn1988antiscreening}
\begin{equation}
    B_\mathrm{crit} = \frac{m_\mathrm{v}^2}{e},
\end{equation}
where there is a classical instability in the magnetic field. In Ref.~\cite{ho2019classical} we demonstrated that this instability leads to monopole production via a classical process, and that at \(B = B_\mathrm{crit}\) the energy barrier to Schwinger production vanishes. We therefore expect that, in the field theory, \(S_\mathrm{inst}\) will vanish at \(B_\mathrm{crit}\).

\section{Numerical Methods} \label{sec:methods}
\subsection{Symmetry of the instanton}
The instanton solution we are searching for is a saddle point of this action with a single negative mode, in a constant, homogeneous background magnetic field. Choosing the magnetic field to point along the \(x_3\) direction, with field strength \(B\), the background U(1) field tensor is
\begin{equation}
    f_{\mu \nu}^\mathrm{ext} = (\delta_{\mu 1} \delta_{\nu 2} - \delta_{\mu 2} \delta_{\nu 1})  B.
\end{equation}
It is clear that this is invariant under rotations in the \(x_3\)-\(x_4\) plane (Euclidean boosts). As a consequence, if instanton solutions exist at all, there must be an instanton solution to the field equations that obeys this symmetry. In weak fields this is the circular worldline instanton identified in Ref.~\cite{affleck1981monopole} and described in the previous section. We proceed by exploiting this symmetry, changing to ``cylindrical'' coordinates
\begin{equation}
    \begin{split}
        x &= x_1, \\
        y &= x_2, \\
        \rho &= \sqrt{x_3^2 + x_4^2}, \\
        \chi &= \arctan(x_4 / x_3).
    \end{split}
\end{equation}

As noted in Ref.~\cite{affleck1981monopole}, the symmetry of the system means that one may choose a gauge such that all fields are independent of \(\chi\), and the gauge field component \(A_{\chi}\) vanishes. Such a field configuration has the action
\begin{equation} \label{eq:3dContinuumAction}
\begin{split}
    S = 2 \pi \int \rho \, \mathrm{d} x \, \mathrm{d} y \, \mathrm{d} \rho \bigg[\frac{1}{2} \tr F_{i j} F^{i j} &+ \tr D_i \Phi D^i \Phi \\
    &+ V(\Phi)\bigg],
\end{split}
\end{equation}

where \(i, j\) represent \(x, y, \rho\). This action is similar to the three dimensional energy density we used in Ref.~\cite{ho2019classical} to study static field configurations, differing only in the Jacobian. We are therefore able to use similar methods to compute the desired instanton solutions, working on a three dimensional lattice with three gauge field components.

\subsection{Lattice Discretisation}
In order to solve the equations of motion that arise from varying \eqref{eq:3dContinuumAction}, we must discretise the action. The symmetry of the problem means that we only need to consider a three dimensional grid of points \(\vec{x} = (n_x, n_y, n_\rho) a\), where \((n_x, n_y, n_\rho)\) are integers and \(a\) is a fixed lattice spacing. Because the coordinate curves of \(x\), \(y\), and \(\rho\) form a Cartesian lattice, the nontrivialities that usually occur when performing a lattice discretisation in curvilinear coordinates are circumvented.

In our lattice theory, the scalar field \(\Phi(\vec{x})\) is defined on lattice sites, whilst the gauge field is defined via link variables \(U_i(\vec{x})\). The discretised form of Eq.~\eqref{eq:3dContinuumAction} is the lattice action

\begin{widetext}
\begin{equation} \label{eq:latticeAction}
    S_\mathrm{lat} = 2 \pi \sum_{\vec{x}} \left\{ \sum_{i < j} \rho^{\square}_{ij}(\vec{x}) \left[ 2 - \tr U_{ij}(\vec{x}) \right] + 2 \sum_i \bar{\rho}_{i}(\vec{x}) \tr \left[ U_i(\vec{x}) \Phi(\vec{x} + \hat{\imath}) U_i^\dagger(\vec{x}) - \Phi(\vec{x}) \right]^2 + \rho(\vec{x}) V(\vec{x}) \right\}.
\end{equation}
\end{widetext}

Here, \(U_{ij}\) is used to denote the standard Wilson plaquette,
\begin{equation}
    U_{ij}(\vec{x}) = U_i(\vec{x}) U_j(\vec{x} + \hat{\imath}) U_i^\dagger(\vec{x} + \hat{\jmath}) U_j^\dagger(\vec{x}).
\end{equation}
We also define appropriately averaged Jacobian factors:
\begin{equation}
\begin{split}
    \rho_{ij}^{\square}(\vec{x}) &= \frac{1}{4} \left[\rho(\vec{x}) + \rho(\vec{x} + \hat{\imath}) + \rho(\vec{x} + \hat{\imath} + \hat{\jmath}) + \rho(\vec{x} + \hat{\jmath})\right], \\
    \bar{\rho}_i(\vec{x}) &= \frac{1}{2} \left[\rho(\vec{x}) + \rho(\vec{x} + \hat{\imath})\right].
\end{split}
\end{equation}
In order to perform calculations it is necessary to impose boundary conditions at \(\rho = 0\) and \(\rho \to \infty\). This is complicated slightly by the fact that the \(U_\rho(\vec{x})\) links are located between lattice points: in our notation the link \(U_\rho(n_x, n_y, n_\rho)\) is located at \((n_x, n_y, n_\rho + \tfrac{1}{2})\). We choose a discretisation such that \(n_\rho\) takes half-integer values in \([-\tfrac{1}{2},R - \tfrac{1}{2}]\), where \(R \in \mathbb{Z}\) is the number of lattice points in the \(\rho\) direction. We then impose boundary conditions at the origin that are compatible with the instanton solution:
\begin{equation}
    \begin{split}
        \Phi(x,y,-\tfrac{1}{2}) &= \Phi(x,y,\tfrac{1}{2}), \\
        U_{x,y}(x,y,-\tfrac{1}{2}) &= U_{x,y}(x,y,\tfrac{1}{2}), \\
        U_\rho(x,y,-\tfrac{1}{2}) &= \mathbb{I}_2;
    \end{split}
\end{equation}
In the continuum limit this is equivalent to imposing symmetry about the origin on \(\Phi\) and \(A_{x,y}\), and imposing antisymmetry about the origin on \(A_\rho\).

At \(n_\rho = R\) we impose reflecting boundary conditions
\begin{equation}
    \begin{split}
        \Phi(x,y,R+\tfrac{1}{2}) &= \Phi(x,y,R-\tfrac{1}{2}), \\
        U_{x,y,\rho}(x,y,R+\tfrac{1}{2}) &= U_{x,y,\rho}(x,y,R-\tfrac{1}{2}).
    \end{split}
\end{equation}
In the \(x\) and \(y\) directions we impose periodic boundary conditions: for \(n_x\) and \(n_y\) taking integer values in \([0,L]\),
\begin{equation}
    \begin{split}
        \Phi(L+1,y,\rho) &= \Phi(0,y,\rho), \\
        U_{x,y,\rho}(L+1,y,\rho) &= U_{x,y,\rho}(0,y,\rho), \\
        \Phi(x,L+1,\rho) &= \Phi(x,0,\rho), \\
        U_{x,y,\rho}(x,L+1,\rho) &= U_{x,y,\rho}(x,0,\rho).
    \end{split}
\end{equation}

After symmetry breaking, the theory retains a U(1) symmetry that defines the electromagnetic field. On the lattice, one can define the operator \cite{davis2000topological}
\begin{equation}
    \Pi_+(\vec{x}) = \frac{1}{2}\left( \mathbb{I}_2 + \frac{\phi(\vec{x})}{|\Phi(\vec{x})|} \right)
\end{equation}
that projects out this subgroup. A U(1) link variable may be defined
\begin{equation}
    u_i(\vec{x}) = \Pi_+(\vec{x}) U_i(\vec{x}) \Pi_+(\vec{x} + \hat{\imath}).
\end{equation}

This may then be used to define an abelian field strength tensor
\begin{equation}
    f_{ij}(\vec{x}) = \frac{2}{e} \mathop{\mathrm{arg}} \tr u_i(\vec{x}) u_j(\vec{x} + \hat{\imath}) u_i^\dagger(\vec{x} + \hat{\jmath}) u_j^\dagger(\vec{x}).
    \vspace{10pt}
\end{equation}
Our goal is to find saddle points of the action~\eqref{eq:latticeAction} with an external magnetic field present. This poses the question of how such an external field can be fixed without adding unphysical terms to the action or equations of motion. The solution to this is to take advantage of the periodic boundary conditions in the \(x\) and \(y\) direction: these quantise the magnetic flux \(\sum_x \sum_y f_{xy}(x,y,\rho)\) to integer multiples of \(4 \pi / e\). Providing the gradient descent updates (described in detail in the next section) are sufficiently small, the deformation of the fields is essentially continuous, so the magnetic flux through the lattice is unchanged under gradient flow (unless a monopole-antimonopole pair is formed or annihilated). We therefore only need to choose initial conditions with the desired magnetic flux to give a solution with the desired external field.

In practice, we impose the unitary gauge \(\Phi(\vec{x}) = \varphi(\vec{x}) \sigma_3\), where \(\{\sigma_1, \sigma_2, \sigma_3\}\) are the Pauli matrices. In this case, \(u_i(\vec{x}) = [U_i(\vec{x})]_{11}\), and a constant magnetic flux through the \(x\)-\(y\) plane may be added to any given initial field configuration by linearly superposing the field
\begin{equation}
\begin{split}
    U_x(\vec{x}) &= \exp(i B y \sigma_3), \\
    U_y(\vec{x}) &= \exp(-i B x \sigma_3).
\end{split}
\end{equation}

\subsection{Saddle point solutions in lattice gauge theory}
The problem of finding saddle point solutions of the action is a far more difficult task than minimisation due to the effects of the negative mode. A na\"{i}ve gradient descent algorithm, for example, will diverge from the desired solution along the negative mode until a local minimum is found.

In previous work~\cite{ho2019classical, ho2020electroweak} we have adapted an algorithm~\cite{chigusa2019bounce} used to find bounce solutions in cosmology. This is an appealing choice because it requires only first order gradient information and is relatively simple to implement. However, it has drawbacks due its nonmonotonic convergence, and the sensitivity of its convergence to the choice of hyperparameters.

In this work, we choose instead to use gradient squared descent: rather than extremise the action \(S_\mathrm{lat}\), we instead choose the objective function
\begin{widetext}
\begin{equation} \label{eq:gradSquared}
    \mathcal{G}^2[\Phi, U_i] = \sum_{\vec{x}} \left[\mathrm{Tr} \left(\frac{1}{\rho(\vec{x})}\frac{\partial S_\mathrm{lat}}{\partial \Phi(\vec{x})}\right)^2 + \sum_j \mathrm{Tr} \left(\frac{i}{\rho(\vec{x})}\frac{\partial S_\mathrm{lat}}{\partial U_j(\vec{x})} \right)^2 \right],
\end{equation}
\end{widetext}
obtained by taking an inner product of the gradient of the action with itself. In the above, the partial derivative with respect to the link variables includes a projection onto the tangent space to SU(2). A similar technique was used to find Electroweak sphaleron configurations on small lattices in Ref.~\cite{hindmarsh1993origin}.

\(\mathcal{G}^2[\Phi, U_i]\) has global minima at all stationary points of the action. These may be identified as saddle points by the fact that they are not minima of the action, and distinguished from spurious local minima of \(\mathcal{G}^2\) by the vanishing of the objective function. Given appropriate initial conditions, a gradient descent algorithm minimising \(\mathcal{G}^2\) will converge on the desired instanton solution. In the case of the Schwinger instanton, suitable initial conditions can be generated from single monopole solutions using our knowledge of the instanton in the weak field limit.

While the algorithm is simple to state, its implementation can be difficult due to the complexity of the objective function \(\mathcal{G}^2\). The full expression for the function is too long to state here, and its efficient calculation can be challenging. To overcome these difficulties, we make use of the automatic differentiation tools available in the TensorFlow library~\cite{tensorflow2015}. These enable computation of the derivatives of arbitrarily complicated functions by algorithmically applying the chain rule to elementary operations. This distinct from numerical differentiation using, for example, finite differences, which introduce a discretisation error. It is also distinct from symbolic differentiation because it stores only the values of intermediate expressions, rather than expressing the derivative as a function. This memoisation can result in a significant improvement in efficiency compared to symbolic differentiation: the computational cost of evaluating the automatic derivative of a function is linearly related to the cost of the function itself~\cite{speelpenning1980compiling}, with a proportionality constant of less than 10 \cite{comsa2019so8}. A review of the use and implementation of automatic differentiation can can be found in Ref.~\cite{baydin2017automatic}.

To find the instanton solutions, we used a gradient descent algorithm with momentum \cite{rumelhart1986learning} to minimise the gradient squared function~\eqref{eq:gradSquared}. To further speed convergence in regions where the gradients are shallow, the gradients of the scalar field and link variables were normalised at each step; the gradient descent update was (omitting the momentum term for brevity)
\begin{widetext}
\begin{equation}
\begin{split}
    \Phi(\vec{x}, \tau + \Delta \tau) &= \Phi(\vec{x}, \tau) - \frac{\Delta \tau}{\sqrt{\sum_{\vec{x}'} [\tfrac{1}{\rho(\vec{x}')}\partial_\Phi \mathcal{G}^2)]^2} + \epsilon} \frac{1}{\rho(\vec{x})}\partial_\Phi (\mathcal{G}^2), \\
    U_i(\vec{x}, \tau + \Delta \tau) &= U_i(\vec{x}, \tau) - \frac{\Delta \tau}{\sqrt{\sum_{\vec{x}', j} [\tfrac{i}{\rho(\vec{x}')}\partial_{U_j} (\mathcal{G}^2)]^2} + \epsilon} \frac{1}{\rho(\vec{x})} \partial_{U_i} (\mathcal{G}^2),
\end{split}
\end{equation}
\end{widetext}
where \(\epsilon\) is a small, positive parameter to avoid divide-by-zero errors and
\begin{equation}
    \begin{split}
        \partial_\Phi(\mathcal{G}^2) &= \frac{\partial(\mathcal{G}^2)}{\partial \Phi(\vec{x})}, \\
        \partial_{U_i}(\mathcal{G}^2) &= \frac{\partial(\mathcal{G}^2)}{\partial U_i(\vec{x})}.
    \end{split}
\end{equation}
The first instanton configurations, in weak fields, were generated using single monopole configurations as the initial condition. Subsequent instanton solutions in stronger fields were generated incrementally by varying the vev and lattice spacing in the theory.

The full code used to find the instanton solutions is publicly available as part of the \texttt{tfmonopoles} Python package~\cite{tfmonopoles2020}.

\section{Instanton for 't~Hooft-Polyakov monopoles} \label{sec:results}

Using the methods described in Section~\ref{sec:methods}, we computed the instanton solution relevant to Schwinger production at magnetic fields strengths up to the critical value, for three values of the boson mass ratio: \(\beta = 0.5\), \(\beta = 1\), and \(\beta = 2\). Our calculations were performed on a \(64^3\) lattice, and an external magnetic flux was fixed by the periodic boundary conditions in the \(x\) and \(y\) direction as described in Section~\ref{sec:methods}. The field strength in units of \(m_\mathrm{v}^2\) was varied by incrementally changing the scalar field vev, while keeping \(\beta\) constant.

\begin{figure*}
    \centering
    \begin{tabular}{c@{}c@{}c@{}}
        \includegraphics[width=0.33\textwidth]{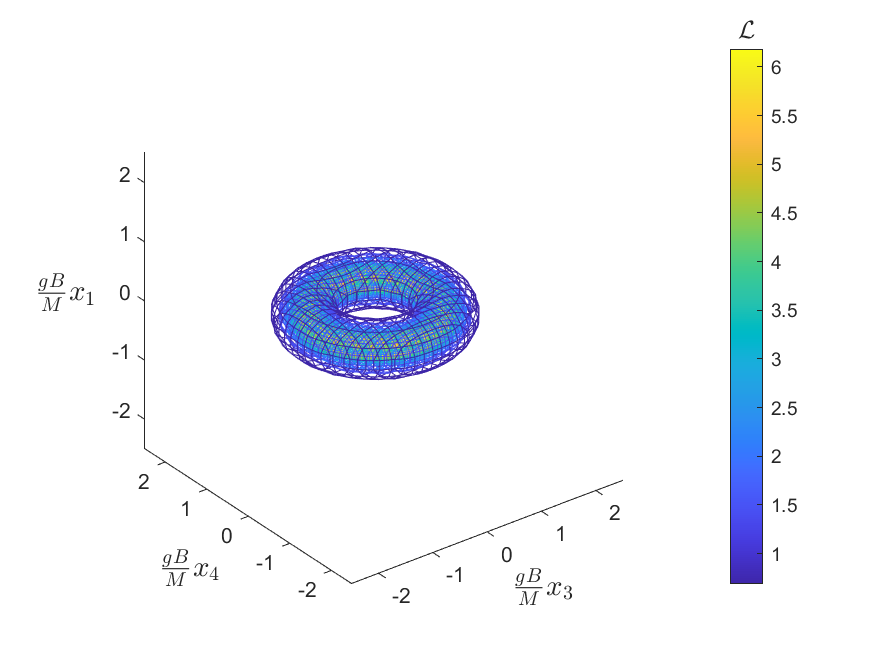} &
        \includegraphics[width=0.33\textwidth]{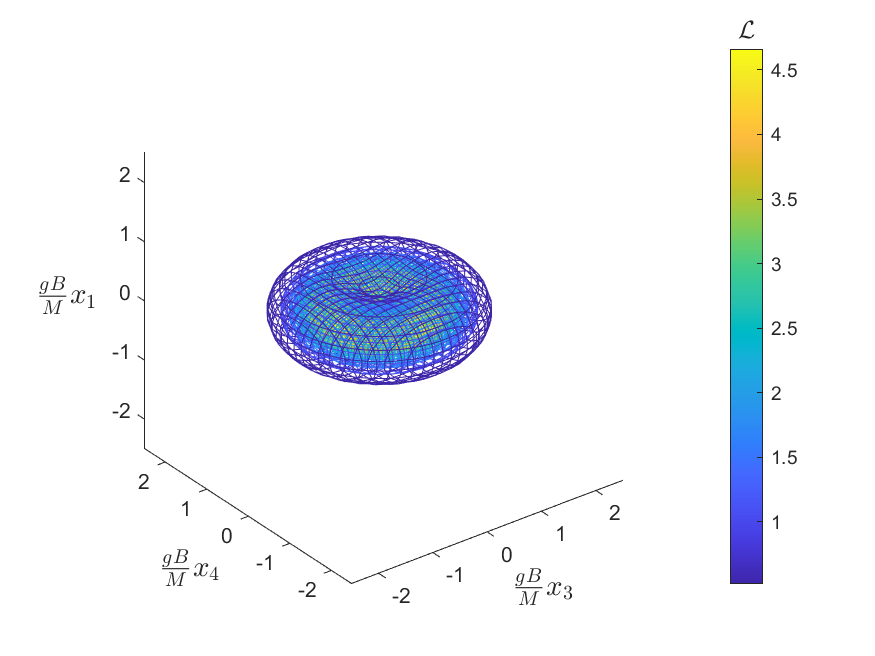} &
        \includegraphics[width=0.33\textwidth]{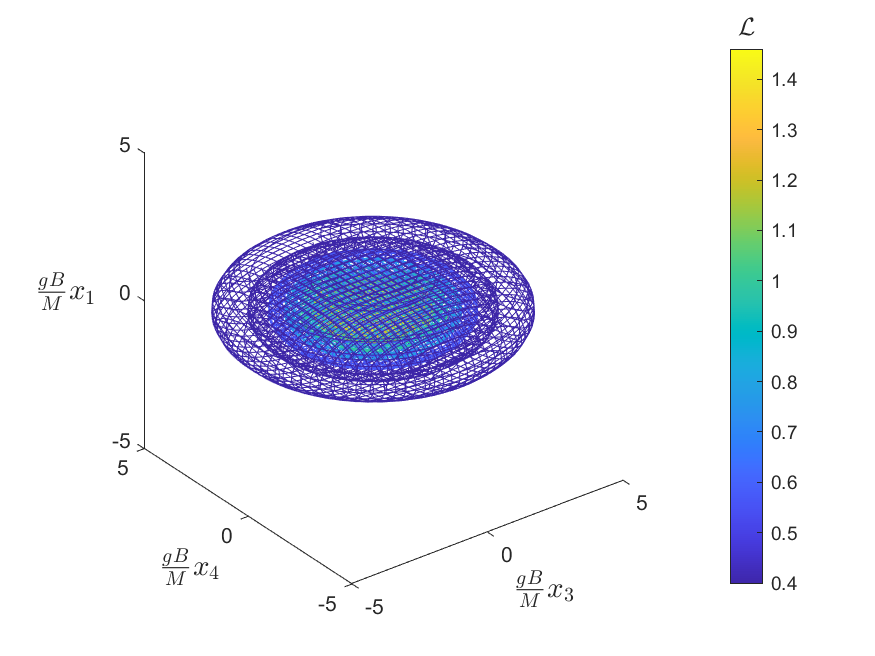} \\
        \includegraphics[width=0.33\textwidth]{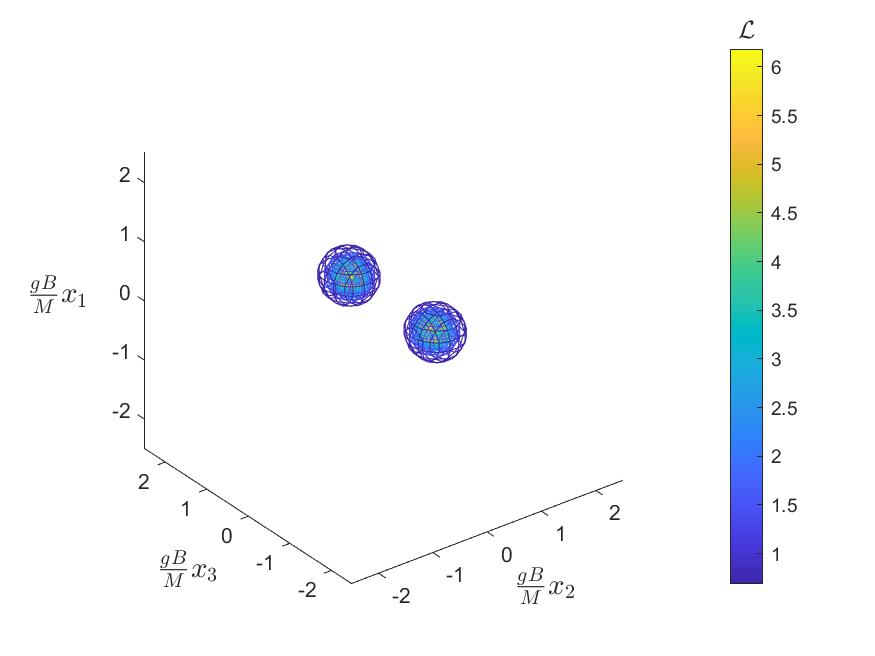} &
        \includegraphics[width=0.33\textwidth]{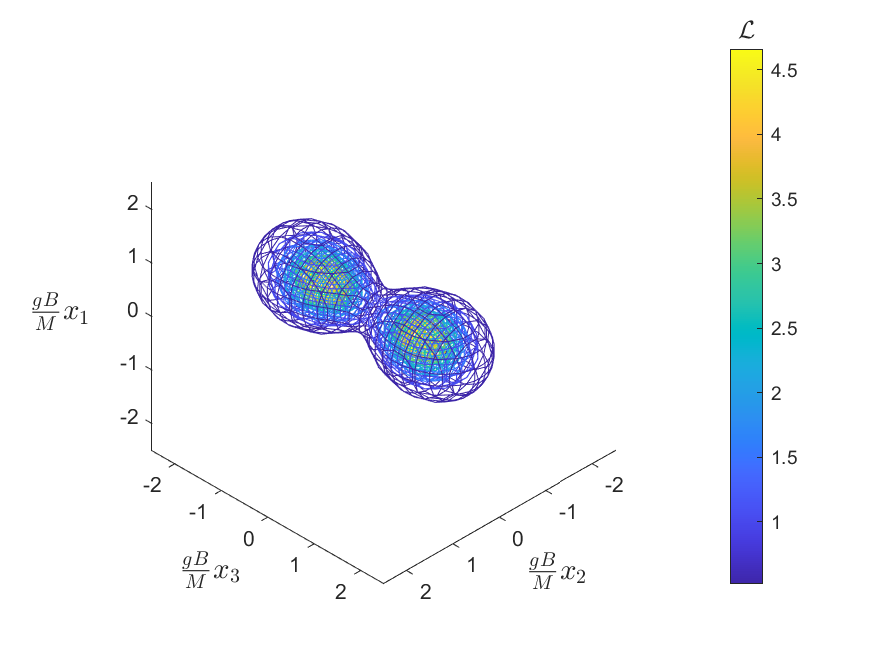} &
        \includegraphics[width=0.33\textwidth]{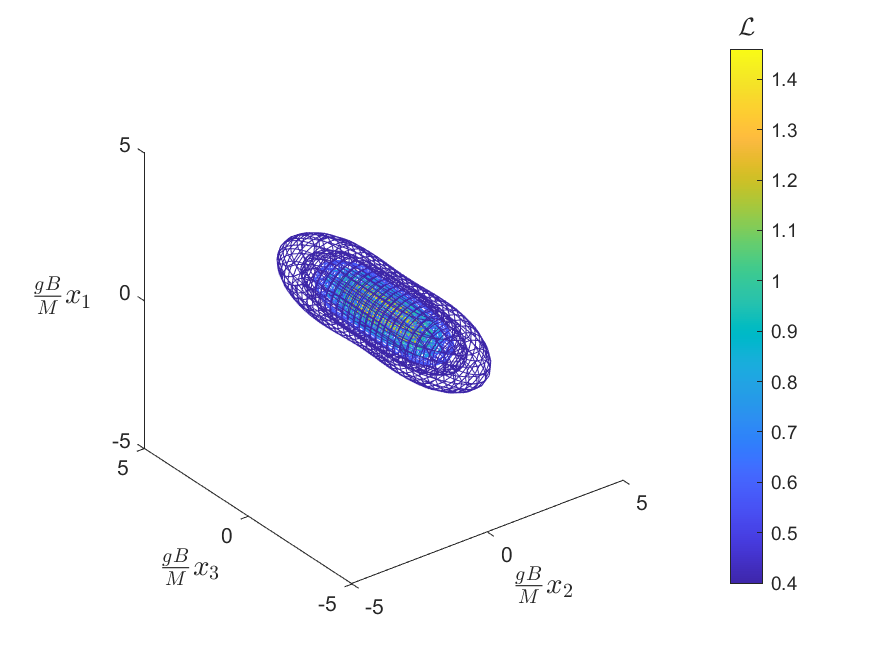} \\
        {(}a{)} \(B/B_\mathrm{crit} = 0.3\) &
        {(}b{) \(B/B_\mathrm{crit} = 0.6\)} &
        {(}c{) \(B/B_\mathrm{crit} = 0.9\)}
    \end{tabular}
    \caption{Lagrangian density contours for instanton solutions for \(\beta = 1\) (\(m_\mathrm{s} = m_\mathrm{v}\)), with the background field subtracted, at different external field strengths. In the upper plots, the $x_2$ dimension is suppressed; in the lower plots, the $x_4$ dimension is suppressed.  Note the difference in scale between the (a), (b) plots and the (c) plots.. Lagrangian density values in units of $m_\mathrm{v}^{-4}$ are shown in the colourbars.}
    \label{fig:contours}
\end{figure*}

\begin{figure*}
    \centering
    \begin{tabular}{c@{}c@{}c@{}}
        \includegraphics[width=0.33\textwidth]{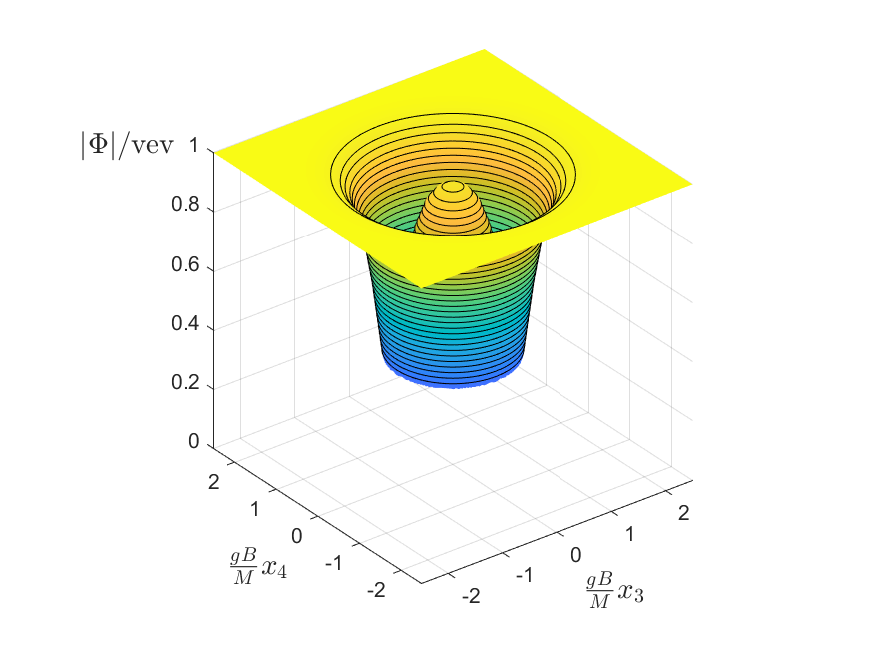} &
        \includegraphics[width=0.33\textwidth]{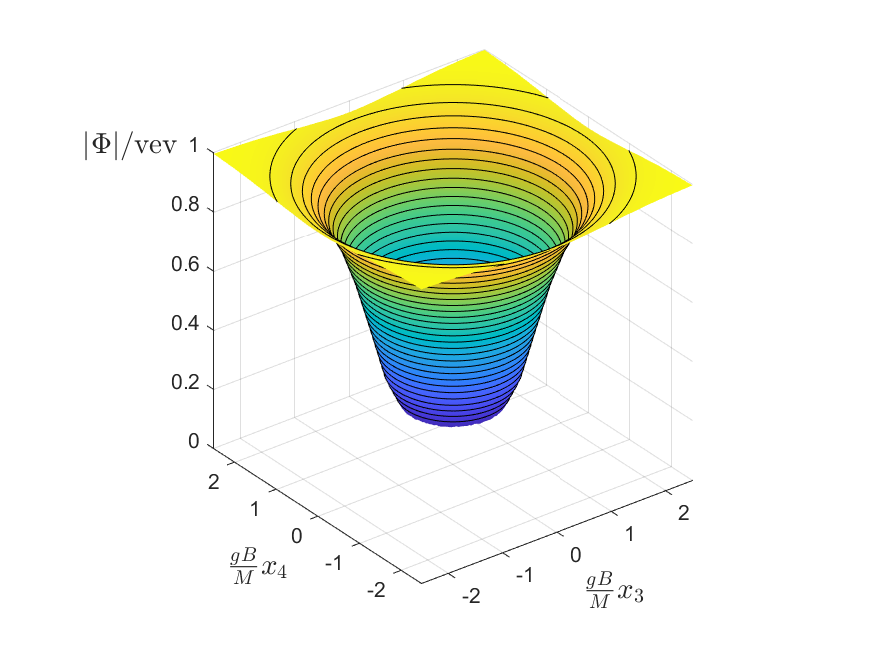} &
        \includegraphics[width=0.33\textwidth]{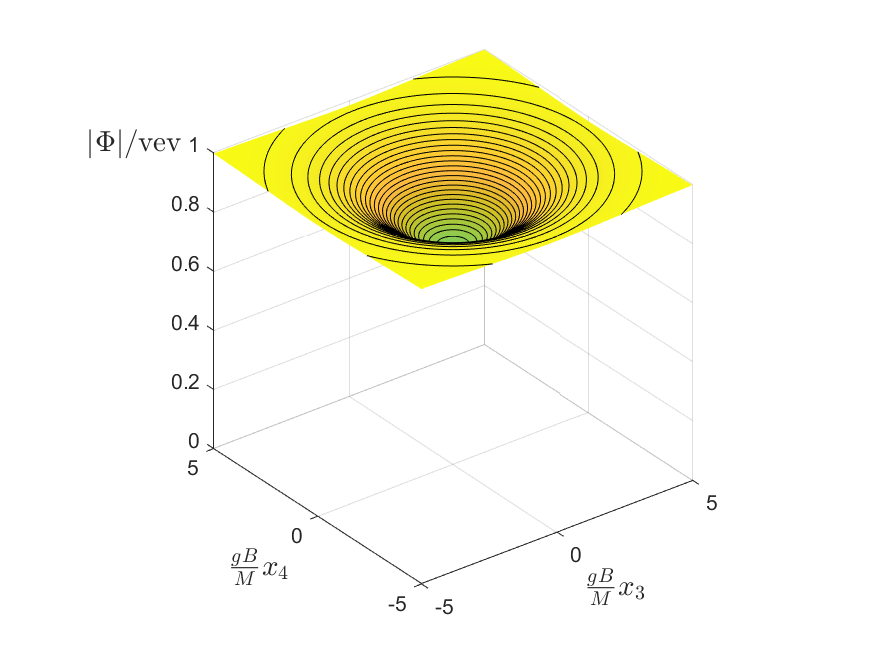} \\
        \includegraphics[width=0.33\textwidth]{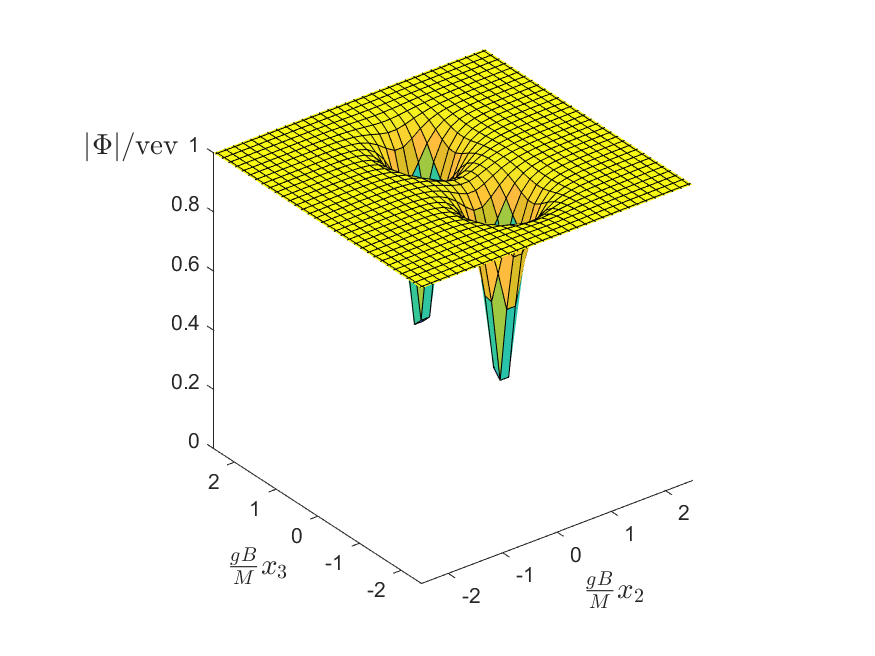} &
        \includegraphics[width=0.33\textwidth]{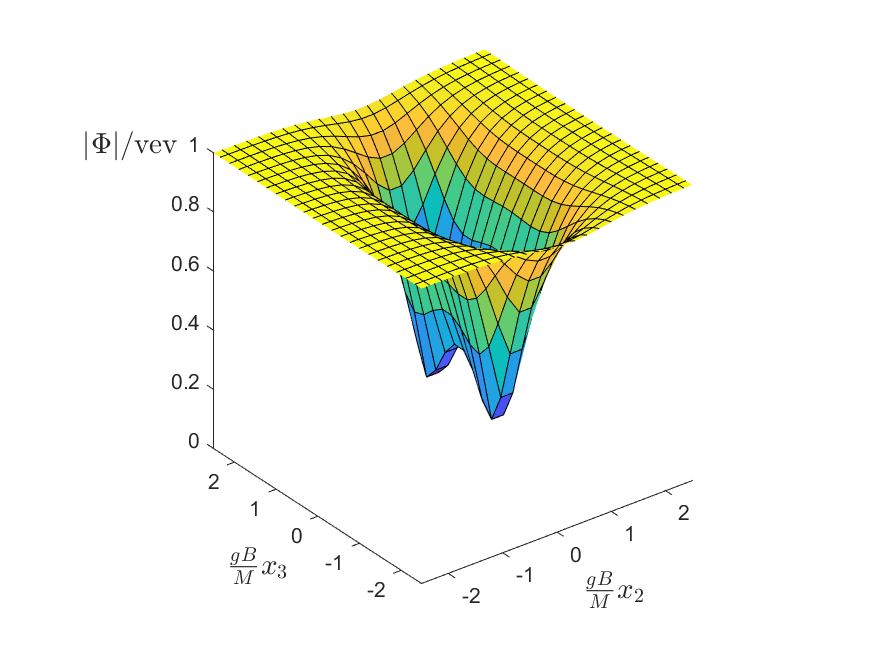} &
        \includegraphics[width=0.33\textwidth]{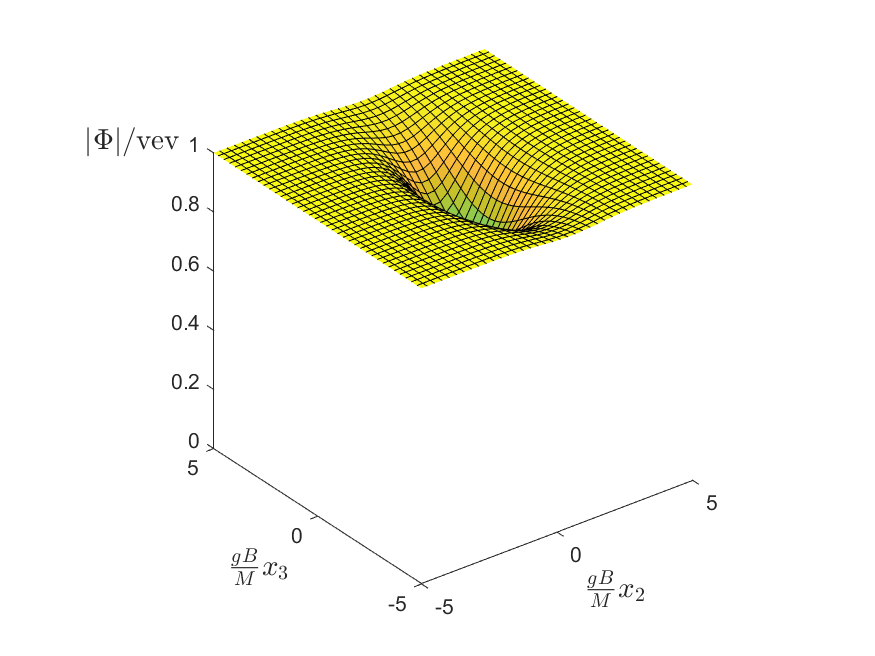} \\
        {(}a{)} \(B/B_\mathrm{crit} = 0.3\) &
        {(}b{) \(B/B_\mathrm{crit} = 0.6\)} &
        {(}c{) \(B/B_\mathrm{crit} = 0.9\)}
    \end{tabular}
    \caption{Surface plots of scalar field magnitude on slices through the instanton centre for different external field strengths, with \(\beta = 1\) (\(m_\mathrm{s} = m_\mathrm{v}\)). Upper plots show the \(x_3\)-\(x_4\) plane, while lower plots show the \(x_2\)-\(x_3\) plane. Note the difference in scale between the (a), (b) plots and the (c) plots.}
    \label{fig:higgsSurfaces}
\end{figure*}

For weak fields the instanton solutions strongly resemble the pointlike approximation; the solution is a circular ring of magnetic charge with localised energy density. An example can be seen in Fig.~\ref{fig:contours}(a) and Fig.~\ref{fig:higgsSurfaces}(a); the energy contours trace a ``doughnut'' shape. The scalar field drops to a minimum on a ring of roughly the worldline instanton radius~\eqref{eq:worldlineInstantonRadius} (in the continuum the scalar field magnitude would vanish), and returns to near the vacuum in the centre of the instanton.

As the field strength increases, the overall extent of the instanton initially stays close to the worldline instanton radius~\eqref{eq:worldlineInstantonRadius}, but the instanton becomes less localised: the hole of the doughnut begins to fill in (see Figs.~\ref{fig:contours}(b) and \ref{fig:higgsSurfaces}(b)). At high external field strengths, the instanton size is significantly larger than the worldline radius (see Figs.~\ref{fig:contours}(c) and \ref{fig:higgsSurfaces}(c)). The minima of the scalar field move closer to the centre of the instanton, until eventually there is a single minimum instead of a ring. At this point, the instanton contains no separated magnetic charges.

As the external field approaches the critical value \(B_\mathrm{crit}\), the scalar field magnitude continuously approaches the vev and the instanton action continuously approaches zero. At the critical field and above, the saddle point solution and the vacuum coincide for all investigated values of \(\beta\).

\begin{figure}[t]
    \centering
    \includegraphics[width=0.5\textwidth]{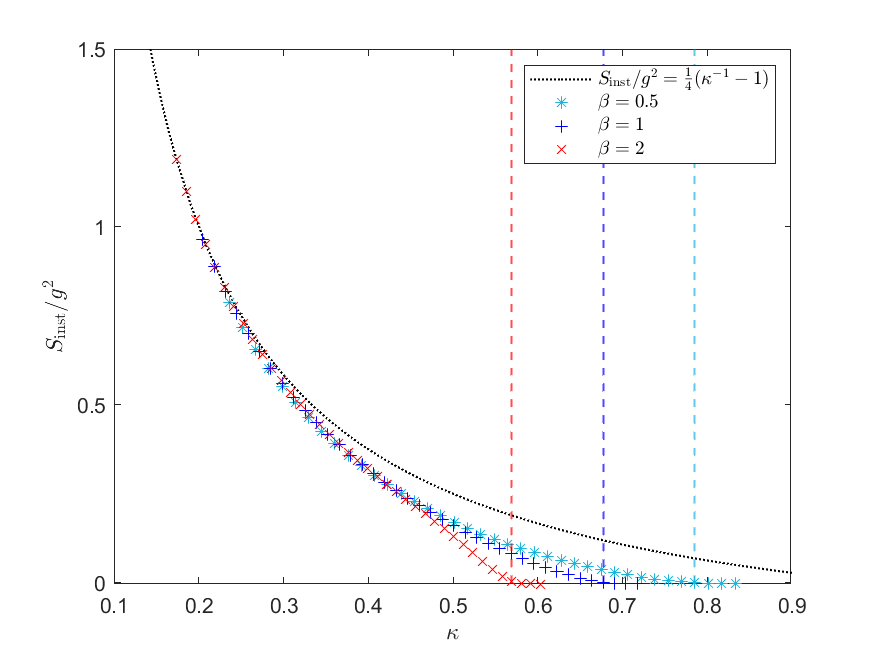}
    \caption{Scaled instanton action plotted against the dimensionless parameter \(\kappa = g^3 B / (4 \pi M^2)\) for different values of \(\beta\). The dotted black curve gives the worldline approximation, and vertical dashed lines indicate the values of \(\kappa\) at which \(B = B_\mathrm{crit}\).}
    \label{fig:kappaPlot}
\end{figure}

To compare the instanton actions it is useful to rewrite the worldline instanton action~\eqref{eq:actionPointlikeLimit} in terms of dimensionless parameters: defining
\begin{equation}
    \kappa = \frac{g^3 B}{4 \pi M^2},
\end{equation}
the action in the worldline approximation is
\begin{equation}
    S_\mathrm{inst} = \frac{g^2}{4}\left(\frac{1}{\kappa} - 1\right).
\end{equation}
The instanton action as a function of \(\kappa\) is plotted in Fig.~\ref{fig:kappaPlot}. For all values of \(\beta\) investigated, the instanton action agrees well with the worldline prediction when \(\kappa\) is small, and plateaus at \(S_\mathrm{inst} = 0\) when \(B = B_\mathrm{crit}\). Note that for different values of \(\beta\), \(B = B_\mathrm{crit}\) corresponds to a different value of \(\kappa\):
\begin{equation}
    \kappa(B_\mathrm{crit}) = \frac{1}{f(\beta)^2},
\end{equation}
where \(f(\beta)\) is defined in Eq.~\eqref{eq:monopoleMass}. Using the results of the high precision calculations in Ref.~\cite{forgacs2005numerical}, this gives \(\kappa(B_\mathrm{crit}) \approx 0.313\) in the limit \(\beta \to \infty\). In the \(\beta \to 0\) limit, \(\kappa(B_\mathrm{crit}) = 1\). Though computing the instanton in these limits is beyond the reach of our current methods, there may be simplifications that render the calculation more tractable in future work, particularly in the BPS limit \(\beta \to 0\), where the 't Hooft--Polyakov monopole solution can be found analytically. 

It is interesting to note that the calculated action diverges from the worldline prediction at \(\kappa \approx 0.3\) for all three values of \(\beta\), but the curves for different values of \(\beta\) remain consistent until \(\kappa \approx 0.5\). This could be because the worldline prediction only accounts for the Coulomb interaction, while 't Hooft--Polyakov monopoles also participate in short-range interactions mediated by the scalar and massive vector bosons. Accounting for these forces could result in a worldline prediction that is accurate at higher values of \(\kappa\), though such a calculation is nontrivial due to nonlocal worldline self interactions.

Another important property shown in Fig.~\ref{fig:kappaPlot} is the fact that the instanton action for 't Hooft--Polyakov monopoles is lower than that for point particles for all values of \(\beta\). This suggests that the finite size effects only enhance monopole pair production rate when compared to the pointlike approximation.

\begin{figure}[b]
    \centering
    \includegraphics[width=0.5\textwidth]{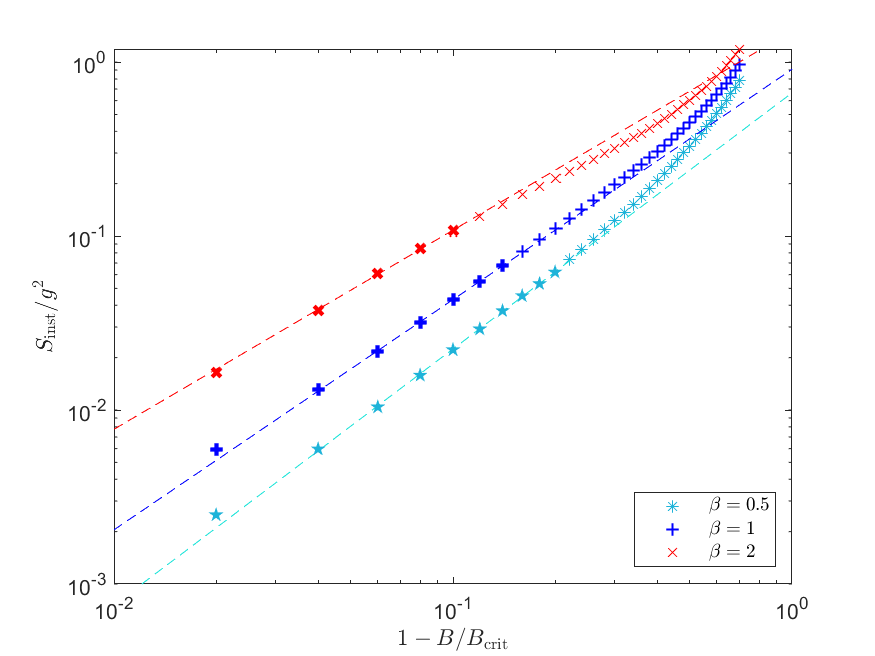}
    \caption{Logarithmic scale plot of scaled instanton action as the external field approaches \(B_\mathrm{crit}\). Dashed lines show fits to the power law regions with exponents given in Table~\ref{tab:exponentTable}. The points used to generate the fits are indicated with filled or thick markers.}
    \label{fig:logPlot}
\end{figure}

The instanton action against \(1 - B / B_\mathrm{crit}\) is plotted on a logarithmic scale in Fig.~\ref{fig:logPlot}. For all three values of \(\beta\), there appears to be power law scaling as the external field approaches its critical value. From the plot the exponents appear to be similar for all three values of \(\beta\), though a numerical fit shows a slight decrease in exponent with increasing \(\beta\). The fitted exponents are given in Table~\ref{tab:exponentTable}.

\begin{table}
    \centering
    \begin{tabular}{c c}
        \hline \hline
        \(\beta\) & Exponent \\ \hline 
        0.5 & 1.47(3) \\
        1 & 1.32(4) \\
        2 & 1.14(6) \\
        \hline \hline
        
    \end{tabular}
    \caption{Exponents \(n\) computed from a numerical fit of the form \(S_\mathrm{inst} = A(1 - B / B_\mathrm{crit})^n\) to the power law regions of the curves in Fig.~\ref{fig:logPlot}. Errors are calculated using the covariance matrix of the least squares linear regression.}
    \label{tab:exponentTable}
\end{table}

\section{Conclusions} \label{sec:conclusion}
We have computed the instanton solution relevant to Schwinger production of 't Hooft--Polyakov monopoles, at external field strengths ranging from the weak field limit to the critical field strength where the instanton energy vanishes. We have confirmed that in the weak field limit, approximating monopoles as point particles with Coulomb interactions gives accurate results. We have also confirmed our earlier result~\cite{ho2019classical} that at the Ambj{\o}rn-Olesen critical field strength \(B_\mathrm{crit} = m_\mathrm{v}^2/e\), monopole production occurs via a classical process.

In Ref.~\cite{ho2019classical}, we noted that the experimental bounds on the mass of heavy charged bosons can be used to place a lower bound on the critical field strength required for unsuppressed monopole production in physical units: this was found to be \(O(10^{23} \ \mathrm{T}) \sim O(10^8 \ \mathrm{GeV^2})\). This bound is unchanged following our most recent analysis, and is far in excess of any known source of magnetic field, so our results are not directly applicable to observation of GUT monopoles in the near future. However, some extensions of the Standard Model predict monopole masses in the TeV range \cite{cho1996monopole, ellis2016price, arunasalm2017electroweak, hung2020topologically}. Extending our calculation to these theories is a promising avenue for future investigation.

Our work is also relevant to the ongoing effort to bound monopole masses using experimental data. If monopoles were sufficiently light, they should be produced via the Schwinger effect in heavy-ion collisions, which generate some of the strongest electromagnetic fields in the Universe \cite{huang2015electromagnetic}. In Ref.~\cite{gould2019schwinger}, we showed that the worldline instanton method is not valid in the spacetime dependent fields of heavy-ion collisions due to effects from the finite monopole size. This calculation shows that finite monopole size effects universally enhance Schwinger production of monopoles, meaning that the predictions from the worldline approximation are suitable for generating lower bounds on monopole masses. A key extension to this calculation, planned for future work, is to recompute the instanton in the spacetime dependent fields present in ultrarelativistic heavy-ion collisions. This will provide the first reliable prediction of the cross section for magnetic monopole production in high energy particle collisions.

\section*{Acknowledgements}
The authors wish to acknowledge the Imperial College Research Computing Service for computational resources. D.L.-J.H would like to thank Megan Wilson for useful discussions and references. D.L.-J.H. was supported by a U.K. Science and Technology Facilities Council studentship. A.R. was supported by the U.K. Science and Technology Facilities Council grants ST/P000762/1 and ST/T000791/1 and Institute for Particle Physics Phenomenology Associateship.

\bibliography{refs}

\end{document}